\documentclass[12pt,preprint]{aastex}
\usepackage{epsfig}
\usepackage{amssymb}
\usepackage[dvips,bookmarks=false]{hyperref}
\newcommand{\degree}{\ensuremath{^\circ}}

\shortauthors{Lattanzi \emph{et al.}}

\begin{document}

\title{Submillimeter Spectrum of Formic Acid}

\author{Valerio Lattanzi\altaffilmark{1,2}, Adam Walters\altaffilmark{2}}

\affil{1. Department of Physics, University \lq \lq La Sapienza'',
P.le Aldo Moro 2, Rome, Italy}

\affil{2. Centre d'Etude Spatiale des Rayonnements, CNRS, Paul Sabatier University, 9, avenue du
Colonel Roche - Boite postale 44346, 31028 Toulouse Cedex 4,
France. valerio.lattanzi@cesr.fr}

\author{Brian J. Drouin\altaffilmark{3}, and John C. Pearson\altaffilmark{3}}

\affil{3. Jet Propulsion Laboratory, California Institute of
Technology, Pasadena, CA 91109-8099}

\begin{abstract}

We have measured new submillimeter-wave data around 600 GHz and around 1.1 THz for the $^{13}$C isotopologue of formic acid and for the two deuterium isotopomers; in each case for both the trans and cis rotamer. For \emph{cis}-DCOOH and \emph{cis}-HCOOD in particular only data up to 50 GHz was previously available. For all species the quality and quantity of molecular parameters has been increased providing new measured frequencies and more precise and reliable frequencies in the range of existing and near-future submillimeter and far-infrared astronomical spectroscopy instruments such as Herschel, SOFIA and ALMA.

\end{abstract}


\section{Introduction}

Formic acid (HCOOH) is the simplest organic acid and the first identified
in the interstellar medium \citep{zuc71}. It has been observed principally in star-forming regions such as Orion KL, Sgr B2, Sgr A and W51
\citep{liu02,liu01,minh92,winneg75} and is associated with hot-cores, and massive star formation. Recently it has also been shown to be present in some hot-corinos (for example that around IRAS 16293-2422) associated
with formation of stars similar to our Sun \citep{botti07,cazaux03}. Several detections in the cold dark cloud L134N/L183 \citep{irvine90,turner99,requena07} have also been reported. Formic acid is also present in chondritic meteorites \citep{briscoe93} and millimeter-wave lines have been observed in outgassing from the Comet Hale-Bopp \citep{bockelee00}. It is a species of intermediate complexity, and as such, may indicate the transition between simpler species that are formed in the gas phase and complex hydrogenated species like CH$_3$OCHO that are believed to form by grain chemistry \citep{turner99}.
The detection of icy HCOOH in star-forming regions \citep[e.g.][]{keane01}
and dense molecular clouds before the onset of star-formation \citep{knez05}
indicates that its formation occurs on grain surfaces during the cold phase
of star-formation but gas-phase formation processes cannot yet be ruled out.
Since it shares common structural elements with biologically important species
such as acetic acid and glycine the study of formic acid should aid future
searches for these molecules and give information on possible reaction paths.

Formic acid has also been observed throughout the Earth's troposphere in liquid,
aerosol, and vapor phases \citep[see for example][]{khare99} and is an important
oxygenated volatile organic compound (OVOC).

Formic acid has two rotameric isomers, defined by whether the two H-atoms are
$\textit{cis}$ or $\textit{trans}$ to each other. Both rotamers are near
prolate asymmetric tops with a planar structure and $C_s$ symmetry
\citep{lerner57,winnem02}. The \emph{cis}-rotamer lies approximately 1365 cm$^{-1}$ higher in energy than the $\textit{trans}$. The latter is hence about 800 times more abundant at room temperature and the $\textit{trans}$ rotamer is assumed if not otherwise specified in the text. All the radioastronomical observations reported above refer to the trans rotamer and the main isotopologue, since, to our knowledge, there are no clear detection of the other species so far. Possible identifications of the other \textit{trans} isotopologues are reported on the NIST web database\footnote{http://physics.nist.gov/PhysRefData/Micro/Html/contents.html} of recommended rest frequencies for observed interstellar molecular microwave transitions.

Many studies have been devoted to $\textit{trans}$-formic acid. The microwave
and millimeter wave rotational spectrum has been studied since the beginning
of the 1950s \citep[for example][]{bellet71_1} and a list of works before 1978
can be found in \citet{wille80}. For HCOOH the components of the dipole
moment along the \emph{a} and \emph{b} axes have values of respectively 1.4071(8) D and 0.227(10) D \citep{weber}. The hyperfine structure due to the interaction with the nuclear
spin of the protons is only a few kHz even for the lowest transitions and
has only been measured at radio-frequencies using a molecular-beam
spectrometer \citep{chardon76}. Several studies have been carried out on the
isotopologues and substitution structures determined \citep{lerner57,bellet71_2,welli80,wille78}.
Work has been done both by FTIR and by microwave spectroscopy on the
vibrational  states \citep[][and references therein]{baska06_1} including the
two lowest $\nu_7$ and $\nu_9$ \citep[for example][]{baska06_2} most pertinent to astrophysical detection.

The $\textit{cis}$ form was first detected by microwave spectroscopy and
the components of the dipole moment determined as $\mu_a$ = 2.65(1) D and
$\mu_b$ = 2.71(1) D \citep{hocking76}. Later, rotational constants of several
isotopologues were determined leading to a substitution $r_s$ structure \citep{bjarnov78}.
The analysis of the rotational spectrum was recently extended to the
submillimeter wave range \citep{baska06_3}.

Recently, detailed ab initio calculations have been carried out for both
the $\textit{cis}$ and the $\textit{trans}$ rotamers \citep{demaison07}.
The quadratic, cubic, and semi-diagonal quartic force fields of $\textit{cis}$-
and $\textit{trans}$-formic acid were calculated using three different
levels of theory. A semi-experimental equilibrium structure was derived
from experimental ground state rotational constants and rovibrational
interaction parameters calculated from the ab initio force field.

The references above concerning the rotational spectrum of formic acid
are limited to the millimeter-wave and microwave regions (below around
300 GHz). With the launch of the Herschel space observatory it is important
to have predictions for astrophysically relevant molecules like formic
acid between 480-1910 GHz (the operating range of the high-resolution
spectrometer HIFI) with accuracies better than a MHz and preferably better
than 100 kHz. Also predictions at higher frequency may be required for
the other lower resolution instruments. Previous work has shown that
extrapolations of spectra from lower frequency laboratory measurements
may not only be inaccurate but also that predicted uncertainties may be
misleadingly smaller than actual shifts \citep[see for example][]{drouin_1,lattanzi07}.
The spatial distribution of complex organic molecules (COMs) like formic acid
and its correlation with grains and temperature should give valuable
clues to the creation and destruction routes of these species. Spatial
resolution generally improves with increasing in frequency. Frequency
coverage of the ALMA international interferometer (operating in the next
decade) will be restricted by atmospheric absorption but measurements
up to 900 GHz should be possible. ALMA will be very sensitive to
the detection of  COMs not only because of the combined detecting area
of up to 80 antennas but also because of its high-spatial resolution
that could be used to focus on regions of high-concentration. Hence
the study of the $\textit{cis}$ isomer and isotopologues of formic acid
would seem feasible. These could be used as probes of the physical
conditions and of formation reactions in the regions where they are
observed. In particular it would be interesting
to see if the overabundance of deuterium isotopologues observed for other
species \citep[see for example][]{parise06} also occurs in formic acid.

The submillimeter-wave spectrum of $\textit{trans}$ and $\textit{cis}$-HCOOH
and $\textit{trans}$-H$^{13}$COOH has recently been measured at Cologne
between 835 - 993 GHz \citep{winnem02} who also report lower resolution far-infrared Fourier transform spectra for the same species between 20-100 cm$^{-1}$ (600-3000 GHz).
The objectives of this work were to measure for the first time high-frequency
rotational transitions of the two deuterated isotopologues DCOOH and HCOOD
and to give at the same time complementary data for H$^{13}$COOH.

\section{Experimental}

The measurements were carried out at JPL using the spectrometer
system developed there and described in detail elsewhere \citep{drouin_2,maiwald05}. 
Briefly, a millimeter-wave module, with 10-100 mW of output power, and a series
of commercial (Virginia Diodes) and JPL built multiplier chains are
used to produce THz radiation. The radiation source is a sweep
synthesizer phase-locked to a frequency standard with a precision of
1 part in $10^{12}$, so the frequency error depends entirely in
determining the line center. Tone burst modulation (100 kHz tone, 0.5 kHz
burst, 0.10-0.15 Volts amplitude) is used as well as a tunable
YIG filter for suppression of spurious harmonics. A double pass of
the microwave radiation through the cell is used to improve
sensitivity. This is achieved using a polarizing beamsplitter at
45\degree \mbox{} to the quasi-optical beam-path between the source and the
cell. The beamsplitter orientation is chosen to transmit a maximum of incident radiation;
a rooftop reflector at the end of the cell rotates the
polarization by 90\degree \mbox{} and reflects the beam back through the cell. The
beamsplitter then reflects the beam into a composite Si bolometer
cooled to 2.1 K with pumped $^{4}$He liquid. The detector response
is passed through a preamplifier, demodulated by a lock-in
amplifier at 0.5 kHz and digitized for storage and treatment by
computer. The commercial (ISOTEC) samples were used at room
temperature and in a static cell filled to 70 mTorr with sample. An air leak limited scanning time and an upper pressure limit of 250 mTorr (sample and air) was used.
The cell was conditioned each time the sample was changed. This was
done by filling the cell with the new sample at high pressure, leaving it for several minutes, pumping it out and repeating the procedure several times. Still significant contamination of previously scanned isotopomers is noticeable in subsequent scans.

\section{Analysis}

Formic acid is a planar molecule with two rotameric configurations,
$\textit{trans}$ and $\textit{cis}$. Scans were performed around 600 and 1200
GHz in regions where major branches of the species overlapped with optimum source power.
For the $\textit{trans}$ species a single scan was in general sufficient for good
signal to noise; for the $\textit{cis}$ species up to ten scans had to be co-added.
The H$^{13}$COOH molecule was measured in the regions 597-607
GHz and 615-650 GHz with some narrower scans in between, for the $\textit{cis}$ isomer.
In the higher range the scans were performed from 1.066 to 1.097 THz. DCOOH was
studied in the ranges 595-650 GHz, 1.075-1.095, 1.100-1.111, 1.122-1.152 and 1.167-1.183 THz.
HCOOD was measured in the same regions with a slightly modified lower
range of 587 to 644 GHz and an additional measurement between 1.187-1.196 THz.
The SPFIT and SPCAT programs \citep{pickett91} were used to fit each
isotopologue separately and to generate improved predictions
that are available on the JPL website.\\
For all our measurements of the $\textit{trans}$ isomers an uncertainty
varying from 50 to 75 kHz was assigned; a 100 kHz uncertainty was used for the $\textit{cis}$ species.
A higher value was given for lines with poor signal to noise or blended transitions.\\
Formic acid is a near prolate symmetric rotor ($\kappa_{\textit{cis}} = -0.96$,
$\kappa_{\textit{trans}} = -0.95$ for the parent species). An $\textit{A}$-reduced Watson Hamiltonian \citep{gordy84} in the $I^r$ representation was used for each isotopologue studied. All the molecular parameters in the fits are defined positively except the quartic.
The exact molecular constants constrained for each species are reported in Tables 1-6.

\subsection{trans-isomers}

For each species we made a final global fit including data from other authors. When the same transition had been remeasured by the present study, our data was used since the estimated uncertainty was in each case smaller. Table \ref{tab:summary1} summarizes the data used for each species including the maximum quantum numbers assigned and the branches with the highest quantum numbers amongst those extensively measured. For example for H$^{13}$COOH the global fit was based on 716 assignments (610 measured) and of these 457 (indicated in the table in parenthesis) are new lines measured in this work. Other lines are taken from work by \citet{welli80} below 185 GHz and from a recent study by \citet{winnem02} in ranges 172-366 GHz and 835-993 GHz.

\begin{figure}
\epsscale{.6}
\plotone{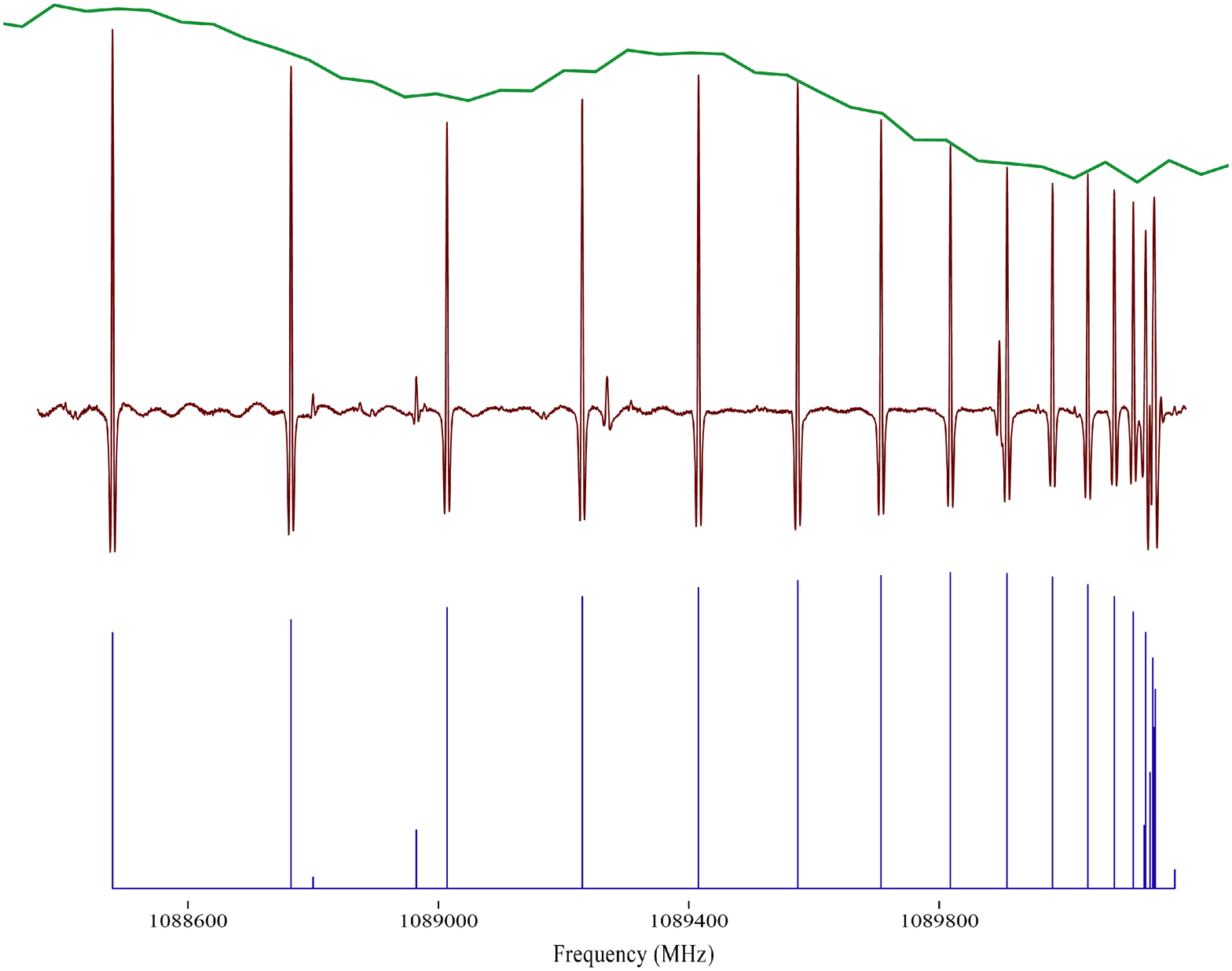}%
\caption{Portion of $^RQ_8(J)$ branch for H$^{13}$COOH (middle).
Prediction of the same branch (bottom) and  power scan in the region (top)
are also reported. \label{fig:trans_h13c}}
\end{figure}

In addition to the new lines measured in the isotopically enriched sample we were able to include some lines identified in natural abundance from spectra taken of the $^{12}$C species between 756-814 GHz. Lines of H$^{13}$COOH, resulting from residual contamination, were also identified during subsequent measurements of DCOOH and as some of these were outside the range of our previous measurements they were also included in the fit. For all subsequent measurements the cell was conditioned to reduce contamination, avoid overlapping spectra and facilitate line identification. Figure \ref{fig:trans_h13c} shows a 1.8 GHz portion of the $^RQ_8(J)$ branch of H$^{13}$COOH (with $J$ from 27 to 12). The corresponding power scan is also shown and can explain small intensity fluctuations. A comparison between the analysis from our work and the most recent published study is reported in Table \ref{tab:trans13}. It can be seen that the higher order centrifugal distortion constants are now better determined. In addition we were able to well constrain the $L_{KKJ}$ and $l_{KJ}$ octic parameters. It should be noted, however, that we were not able to constrain the $L_K$ parameter from the new dataset even without inclusion of the two new octic parameters specified above. The choice of parameters was that which gave the best fit. The new parameter choice is needed to fit the high K transitions, when these are not included $L_K$ may be fitted as previously.

In the analysis of DCOOH, data from \citet{bellet71_1} and \citet{baska96} were included. Lines from \citet{bellet71_1} and \citet{baska99_1} were added to our HCOOD dataset. As can be seen from Tables 1-3 the same set of parameters were used to fit all the \emph{trans} species. For both deuterated species all centrifugal distortion parameters are better constrained and the three octic constants have been determined for the first time. For DCOOH, $\phi_{JK}$ has been constrained for the first time and an additional significant figure has been determined for the other sextic parameters. For HCOOD significant changes in $\phi_{JK}$ and $\phi_K$ should be noted with $\phi_{JK}$ now taking a value much closer to that of DCOOH and HCOOH. For HCOOD an additional significant figure has been determined for  $\phi_J$ and $\phi_{JK}$ has the uncertainty reduced by 7.

\subsection{cis-isomers}

\begin{figure}
\epsscale{.6}
\plotone{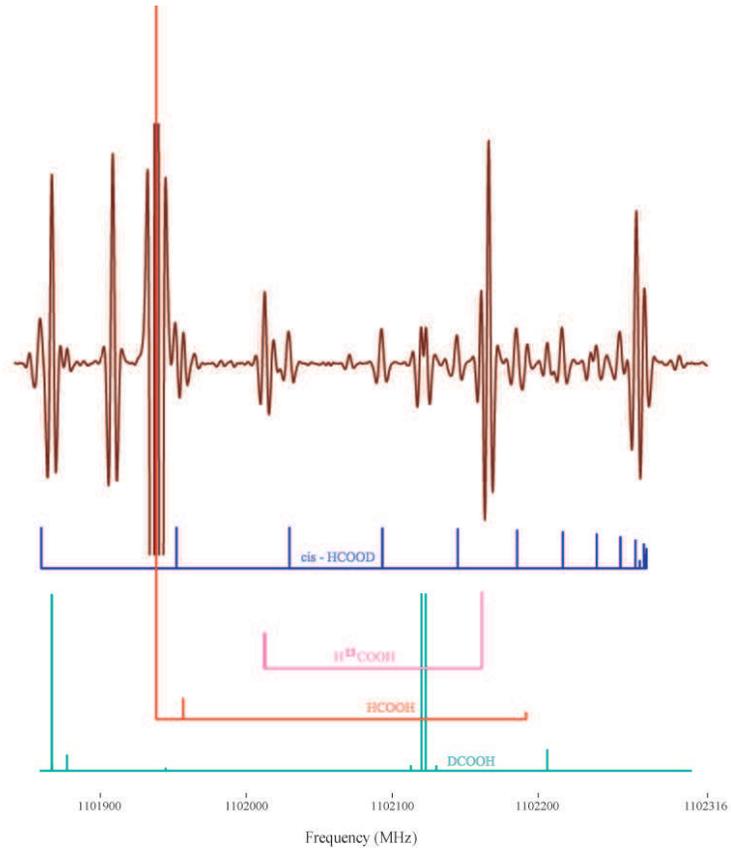}
\caption{HCOOD 400 MHz scan near 1.1 THz. The simulations of the different
isotopologues, present in the spectrum as contamination, are also shown. \label{fig:cis_hcood}}
\end{figure}

Data used in the fits for the \emph{cis} isomers are also summarized in Table \ref{tab:summary2}. The data for the ground state of \textit{cis}-H$^{13}$COOH, previously reported, consisted in 100 lines  between 152 and 375 GHz from \citet{baska06_3} and 28 microwave transitions below 47 GHz from \citet{bjarnov78}. This data was merged with our 96 new assignments for a global analysis, the parameters of which are given in Table \ref{tab:cish13}. The values of the parameters from the two different data sets are generally in good agreement for the quadratic and quartic portions of the Hamiltonian (3$\sigma$ or less except $\delta_K$ for HCOOD).
Three sextic parameters, $\Phi_J$, $\Phi_K$ and $\phi_K$ have been constrained for the first time and a fourth, sextic term, $\Phi_{KJ}$ is determined with an additional significant figure.  Inclusion of the sextic constant $\Phi_J$ improves the fit even though it is determined with an uncertainty of only 5$\sigma$.\\
For both the deuterated species the previous measurements of the ground state have been reported by \citet{bjarnov78}.
Respectively 31 and 24 lines for \textit{cis}-DCOOH and \textit{cis}-HCOOD below 50 GHz were added to our measurements giving more than 110 lines up to 1.2 THz for each isotopomer. A comparison of the molecular parameters obtained with the previous analysis using the low frequency data is given in Tables \ref{tab:cisdcooh} and \ref{tab:cishcood}. There is a good global agreement between rotational and quartic centrifugal parameters although the latter are determined somewhat outside previous uncertainty boundaries. As is to be expected all parameters are better determined. 
For both isotopomers sextic constants have been determined for the first time and for cis-HCOOD the inclusion of two octic parameters improved the fit. In both cases our work improves the number and quality of determinable parameters.\\
An extract of a few hundred MHz of HCOOD spectrum is shown in Figure \ref{fig:cis_hcood}. Simulations of the different isotopologues are overlapped and it can be seen how the contaminating spectra of the species used for previous measurements, are sometimes more intense than \textit{cis} isomer transitions in spite of conditioning of the cell. In spite of this the high signal to noise ratio (S/N) allowed the assignment of almost all
the $^RQ_7(J)$ lines to be made.

\section{Discussion}



Recently, several publications, mostly by Baskakov (see for example 1999a, 1999b, 2003) on formic acid (including the isotopologues and the \emph{cis} rotamer) have been devoted to its excited vibrational states, including infrared data and microwave and millimeter wave measurements of rotational transitions of vibrationally excited molecules. Some of these publications have given analyses both for the vibrational state considered and for the vibrational ground state using combination differences. Our work concerns uniquely the vibrational ground state and focuses on submillimeter and terahertz data of $^{13}$C and deuterium isotopologues in order to provide accurate measured and predicted frequencies for interpretation of far-infrared spectra taken by near future instruments such as HIFI/Herschel, SOFIA and ALMA. 

In particular we have provided new high-frequency rotational data for both rotamers of the two monodeuterated isotopomers DCOOH and HCOOD. For the \emph{trans} rotamers only millimeter wave data below 335 GHz was previously available \citep{baska96,bellet71_1}. The new data allow the determination of three octic parameters and improves other higher order centrifugal distortion parameters. For each \emph{cis} species only one set of measurements below 50 GHz \citep{bjarnov78} has previously been published. For \emph{cis}-DCOOH three sextic parameters and for \emph{cis}-HCOOD five sextic and two octic parameters have been constrained for the first time. Also other centrifugal parameters are better determined. Similarly for \emph{cis}-H$^{13}$COOH only measurements below 370 GHz \citep{baska06_3} have previously been reported and two sextic and one octic parameters have been newly constrained. For \emph{trans}-H$^{13}$COOH \citet{winnem02} recently published data to almost a THz; we have increased the dataset filling in the 600 GHz region and increasing the maximum measured frequency to almost 1.1 THz. The higher order centrifugal constants have been improved.

It is also interesting to review for formic acid the previous state of two major databases used by astronomers for interpreting rotational spectra; that of the JPL\footnote{http://www.spec.jpl.nasa.gov} and of Cologne University (CDMS\footnote{http://www.ph1.uni-koeln.de/vorhersagen/}). These databases contained no entry for any of the \emph{cis} species except the parent \emph{cis}-HCOOH (CDMS). For \emph{trans}-DCOOH and \emph{trans}-HCOOD only an entry in JPL from 1980 existed before this work and all predictions were based on measurements up to only 160 GHz and 237 GHz respectively. For \emph{trans}-H$^{13}$COOH a recent entry in the CDMS catalogue is based on all measurements except the new ones reported here; the new entry in JPL should be slightly more precise.

To summarize the present work provides new experimentally determined frequencies in the THz region. The improvement in the determination of the higher order centrifugal parameters should make predictions of all lines more precise and reliable especially at the higher frequencies of the Herschel operating range. All new predictions are made available on the JPL website.

\section{Acknowledgments}

VL is a student financed by the Universit\'e Franco-Italienne. His
visit to JPL for the measurements was financed by the Observatoire
Midi-Pyr\'en\'ees and the PCMI. A summer project student at Toulouse, Marion Cardon, carried out some of the 
analysis under the supervision of the authors. A portion of this research
was performed at the Jet Propulsion Laboratory, California Institute of Technology,
under contract with National Aeronautics and Space Administration.
Any opinions, findings, and conclusions or recommendations
expressed in this material are those of the author(s) and do not
necessarily reflect the views of the NASA.


\clearpage

\clearpage

\clearpage

\begin{table}[ht]
\begin{center}
\caption{One standard deviation in units of the last decimal place is given
in parentheses.
$^*$Baskakov et al. 2006b.\label{tab:trans13}}
\begin{tabular}{lr@{.}lr@{.}l}
\multicolumn{5}{c}{\textbf{\emph{trans}-H$^{13}$COOH}}\\
\textsc{Constant} & \multicolumn{2}{c}{\textsc{Our analysis}} &
\multicolumn{2}{c}{\textsc{Previous$^*$}}\\
\hline \hline
$A$/MHz                       &   75580&8709(20)     &    75580&8205(20)     \\
$B$/MHz                       &   12053&56814(17)    &    12053&56907(17)    \\
$C$/MHz                       &   10378&99931(16)    &    10378&99871(16)    \\
$\Delta_J$/kHz                &     9&92780(13)      &        9&92604(22)    \\
$\Delta_{JK}$/kHz             &    -84&7656(30)      &      -84&7022(22)     \\
$\Delta_K$/kHz                &   1672&824(37)       &     1670&588(24)      \\
$\delta_J$/kHz                &      1&982195(51)    &        1&982587(92)   \\
$\delta_K$/kHz                &      41&9738(46)     &       41&996(19)      \\
$\Phi_J$/Hz                   &       0&012327(31)   &        0&011898(79)   \\
$\Phi_{JK}$/Hz                &       0&1461(46)     &        0&0978(81)     \\
$\Phi_{KJ}$/Hz                &     -10&617(16)      &      -9&871(28)       \\
$\Phi_K$/Hz                   &     116&65(24)       &      107&90(58)       \\
$\phi_J$/Hz                   &     0&005819(14)     &         0&005865(31)  \\
$\phi_{JK}$/Hz                &     0&0785(22)       &         0&0982(64)    \\
$\phi_K$/Hz                   &    16&51(20)         &        14&10(40)      \\
$L_{JJK}$/mHz                 &    -0&01087(69)      &        -0&00431(37)    \\
$L_{KKJ}$/mHz                 &     0&958(13)        &  \multicolumn{2}{c}{-} \\
$l_{KJ}$/mHz                  &    -0&496(29)        &  \multicolumn{2}{c}{-} \\
$L_{K}$/mHz                   &\multicolumn{2}{c}{-} &         2&501(35)      \\
\hline
\end{tabular}
\end{center}
\end{table}

\clearpage

\begin{table}[ht]
\begin{center}
\caption{One standard deviation in units of the last decimal place is given
in parentheses.
$^*$Baskakov 1996.\label{tab:transdcooh}}
\begin{tabular}{lr@{.}lr@{.}l}
\multicolumn{5}{c}{\textbf{\emph{trans}-DCOOH}}\\
\textsc{Constant} & \multicolumn{2}{c}{\textsc{Our analysis}} &
\multicolumn{2}{c}{\textsc{Previous$^*$}}\\
\hline \hline
$A$/MHz                       &   57709&2246(10)     &   57709&2359(16)    \\
$B$/MHz                       &   12055&97980(23)    &   12055&98297(40)   \\
$C$/MHz                       &    9955&61071(24)    &    9955&61211(41)   \\
$\Delta_J$/kHz                &       9&44044(17)    &       9&44049(65)   \\
$\Delta_{JK}$/kHz             &     -39&5883(15)     &     -39&5749(40)    \\
$\Delta_K$/kHz                &     757&705(13)      &     757&707(44)     \\
$\delta_J$/kHz                &       2&226002(33)   &       2&22726(28)   \\
$\delta_K$/kHz                &      37&5636(22)     &      37&510(17)     \\
$\Phi_J$/Hz                   &       0&012078(39)   &       0&01064(25)   \\
$\Phi_{JK}$/Hz                &       0&1916(15)     &       0&244(17)     \\
$\Phi_{KJ}$/Hz                &      -4&5734(47)     &      -4&639(57)     \\
$\Phi_K$/Hz                   &      33&384(48)      &      33&21(25)      \\
$\phi_J$/Hz                   &       0&0059522(75)  &       0&00708(11)   \\
$\phi_{JK}$/Hz                &       0&10299(72)    & \multicolumn{2}{c}{-}  \\
$\phi_K$/Hz                   &       7&926(40)      &       9&57(47)      \\
$L_{JJK}$/mHz                 &      -0&00558(22)    & \multicolumn{2}{c}{-}  \\
$L_{KKJ}$/mHz                 &       0&2427(21)     & \multicolumn{2}{c}{-}  \\
$l_{KJ}$/mHz                  &      -0&0971(57)     & \multicolumn{2}{c}{-}  \\
\hline
\end{tabular}
\end{center}
\end{table}

\clearpage

\begin{table}[ht]
\begin{center}
\caption{One standard deviation in units of the last decimal place is given
in parentheses.
$^*$Baskakov et al. 1999a.\label{tab:transhcood}}
\begin{tabular}{lr@{.}lr@{.}l}
\multicolumn{5}{c}{\textbf{\emph{trans}-HCOOD}}\\
\textsc{Constant} & \multicolumn{2}{c}{\textsc{Our analysis}} &
\multicolumn{2}{c}{\textsc{Previous$^*$}}\\
\hline \hline
$A$/MHz                       &   66099&4233(17)     &   66099&4337(20)    \\
$B$/MHz                       &   11762&55683(29)    &   11762&55737(29)   \\
$C$/MHz                       &    9969&96440(30)    &    9969&96432(28)   \\
$\Delta_J$/kHz                &      10&20172(22)    &      10&19714(60)   \\
$\Delta_{JK}$/kHz             &     -59&2283(25)     &     -59&2345(33)    \\
$\Delta_K$/kHz                &     989&3958(29)     &     989&421(27)     \\
$\delta_J$/kHz                &       2&153715(40)   &       2&15391(18)   \\
$\delta_K$/kHz                &      43&4084(37)     &      43&393(11)     \\
$\Phi_J$/Hz                   &       0&014245(41)   &       0&01049(39)   \\
$\Phi_{JK}$/Hz                &       0&2244(39)     &  \multicolumn{2}{c}{-}\\
$\Phi_{KJ}$/Hz                &      -6&285(17)      &      -5&459(17)     \\
$\Phi_K$/Hz                   &      50&27(15)       &      49&63(13)      \\
$\phi_J$/Hz                   &       0&006393(10)   &       0&00602(13)   \\
$\phi_{JK}$/Hz                &       0&1043(14)     &       0&2147(96)    \\
$\phi_K$/Hz                   &      12&92(13)       &       3&73(10)      \\
$L_{JJK}$/mHz                 &       0&431(22)      & \multicolumn{2}{c}{-} \\
$L_{KKJ}$/mHz                 &      -0&01126(56)    & \multicolumn{2}{c}{-} \\
$l_{KJ}$/mHz                  &      -0&413(20)      & \multicolumn{2}{c}{-} \\
\hline
\end{tabular}
\end{center}
\end{table}

\clearpage

\begin{table}[ht]
\begin{center}
\caption{One standard deviation in units of the last decimal place is given
in parentheses.
$^a$Baskakov et al. 2006c.\newline
$^b$Fixed at value for the ground state of \textit{cis}-HCOOH.\label{tab:cish13}}
\begin{tabular}{lr@{.}lr@{.}l}
\multicolumn{5}{c}{\textbf{\emph{cis}-H$^{13}$COOH}}\\
\textsc{Constant} & \multicolumn{2}{c}{\textsc{Our analysis}} &
\multicolumn{2}{c}{\textsc{Previous$^a$}}\\
\hline \hline
$A$/MHz                       &   84201&7591(41)     &    84201&7544(67)  \\
$B$/MHz                       &   11687&51692(76)    &    11687&51738(73) \\
$C$/MHz                       &   10249&65418(66)    &    10249&65471(65) \\
$\Delta_J$/kHz                &     8&2900(18)       &       8&2918(12)   \\
$\Delta_{JK}$/kHz             &    -69&8443(97)      &     -69&858(17)   \\
$\Delta_K$/kHz                &   2311&71(20)        &    2311&35(20)    \\
$\delta_J$/kHz                &      1&44685(51)     &       1&44669(52)  \\
$\delta_K$/kHz                &      40&477(20)      &       40&486(23)  \\
$\Phi_J$/Hz                   &       0&0056(11)     &        0&008366{$^b$} \\
$\Phi_{KJ}$/Hz                &     -10&673(31)      &      -11&37(47)  \\
$\Phi_K$/Hz                   &     177&5(21)        &      181&70{$^b$} \\
$\phi_J$/Hz                   &     0&00441(34)      &        0&00409(48)  \\
$\phi_K$/Hz                   &    20&44(41)         &       18&10{$^b$} \\
\hline
\end{tabular}
\end{center}
\end{table}

\clearpage

\begin{table}[ht]
\begin{center}
\caption{One standard deviation in units of the last decimal place is given
in parentheses.
$^*$Bjarnov 1978.\label{tab:cisdcooh}}
\begin{tabular}{lr@{.}lr@{.}l}
\multicolumn{5}{c}{\textbf{\emph{cis}-DCOOH}}\\
\textsc{Constant} & \multicolumn{2}{c}{\textsc{Our analysis}} &
\multicolumn{2}{c}{\textsc{Previous$^*$}}\\
\hline \hline
$A$/MHz                       &   62653&4524(66)    & 62653&4395(87)  \\
$B$/MHz                       &   11690&16785(83)    & 11690&1692(18)  \\
$C$/MHz                       &    9837&91632(65)    &  9837&9145(18)   \\
$\Delta_J$/kHz                &       7&86216(45)   &     7&899(29)   \\
$\Delta_{JK}$/kHz             &     -23&636(12)     &   -22&94(48)    \\
$\Delta_K$/kHz                &     957&18(11)      &   954&4(11)      \\
$\delta_J$/kHz                &       1&65774(24)   &     1&6633(34)  \\
$\delta_K$/kHz                &      37&457(36)     &    37&541(94)   \\
$\Phi_{JK}$/Hz                &       0&0989(60)    &  \multicolumn{2}{c}{-}\\
$\Phi_{KJ}$/Hz                &      -3&522(56)     &  \multicolumn{2}{c}{-}\\
$\Phi_K$/Hz                   &      43&78(58)      &  \multicolumn{2}{c}{-}\\
\hline
\end{tabular}
\end{center}
\end{table}

\clearpage

\begin{table}[ht]
\begin{center}
\caption{One standard deviation in units of the last decimal place is given
in parentheses.
$^*$Bjarnov 1978.\label{tab:cishcood}}
\begin{tabular}{lr@{.}lr@{.}l}
\multicolumn{5}{c}{\textbf{\emph{cis}-HCOOD}}\\
\textsc{Constant} & \multicolumn{2}{c}{\textsc{Our analysis}} &
\multicolumn{2}{c}{\textsc{Previous$^*$}}\\
\hline \hline
$A$/MHz                       &   83962&816(16)    &   83962&785(22)       \\
$B$/MHz                       &   10883&9446(17)   &   10883&9413(29)      \\
$C$/MHz                       &    9624&9474(18)   &    9624&9421(29)      \\
$\Delta_J$/kHz                &       6&6876(18)   &       6&733(36)       \\
$\Delta_{JK}$/kHz             &     -49&378(41)    &     -48&11(81)        \\
$\Delta_K$/kHz                &    1993&69(43)     &    1984&4(48)         \\
$\delta_J$/kHz                &       1&06150(44)  &       1&0628(21)      \\
$\delta_K$/kHz                &      33&567(54)    &      33&911(68)       \\
$\Phi_{J}$/Hz                 &      14&95(73)     & \multicolumn{2}{c}{-} \\
$\Phi_{KJ}$/Hz                &      -3&06(52)     & \multicolumn{2}{c}{-} \\
$\Phi_K$/Hz                   &     118&2(42)      & \multicolumn{2}{c}{-} \\
$\phi_{J}$/Hz                 &       0&00338(28)  & \multicolumn{2}{c}{-} \\
$\phi_K$/Hz                   &     -25&2(31)      & \multicolumn{2}{c}{-} \\
$L_{JK}$/mHz                 &      -3&06(32)     & \multicolumn{2}{c}{-} \\
$l_{KJ}$/mHz                 &      19&8(13)      & \multicolumn{2}{c}{-} \\
\hline
\end{tabular}
\end{center}
\end{table}

\clearpage

\begin{table}[ht]
\begin{center}
\begin{tabular}{ccrr}
\multicolumn{4}{c}{$\textit{trans}$ Species}\\
Isotopologues & Major Branches  &   \multicolumn{1}{c}{Fit} & Other lines \\
\hline
\hline

H$^{13}$COOH & $^{R}Q_8(J_{max}=47$)    & $\sigma$ = 45 kHz     & $J_{max}$ = 56\\
             & $^{Q}R_{K_{max=17}}$(50) & $\sigma_{red}$ = 0.70 & $K_{max}$ = 25\\
             &                          & N = 716(457)         &               \\
\hline
DCOOH        & $^{R}Q_{12}(J_{max}=48$) & $\sigma$ = 84 kHz     & $J_{max}$ = 66\\
             & $^{Q}R_{K_{max=20}}$(52) & $\sigma_{red}$ = 0.98 & $K_{max}$ = 32\\
             &                          & N = 738(537)         &               \\
\hline
HCOOD        & $^{R}Q_{10}(J_{max}=47$) & $\sigma$ = 87 kHz     & $J_{max}$ = 59\\
             & $^{Q}R_{K_{max=20}}$(50) & $\sigma_{red}$ = 0.90 & $K_{max}$ = 22\\
             &                          & N = 641(513)         &               \\
\hline

\end{tabular}
\vspace{1cm}
\caption{Branches with several transitions assigned and with the highest
quantum numbers are reported here. The notation $^{\Delta K_a}\Delta J_{K_a}(J_{low})$ was used. Other lines,
eventually with higher quantum numbers, were used in the fit and the highest $J$ and $K_a$ are reported.
The fit column gives the standard deviation, the reduced error and the numbers of lines used for the
global analysis. In brackets the numbers of lines assigned from our spectra. \label{tab:summary1}
}
\end{center}
\end{table}

\clearpage

\begin{table}[ht]
\begin{center}
\begin{tabular}{ccrr}
\multicolumn{4}{c}{$\textit{cis}$ Species}\\
Isotopologues & Major Branches  &   \multicolumn{1}{c}{Fit} & Other lines \\
\hline
\hline

H$^{13}$COOH & $^{R}Q_7(J_{max}=39$)    & $\sigma$ = 54 kHz     & $J_{max}$ = 39\\
             & $^{Q}R_{K_{max=15}}$(28) & $\sigma_{red}$ = 0.92 & $K_{max}$ = 17\\
             &                          & N = 224(96)          &               \\
\hline
DCOOH        & $^{R}Q_{10}(J_{max}=55$) & $\sigma$ = 67 kHz     & $J_{max}$ = 55\\
             & $^{Q}R_{K_{max=11}}$(27) & $\sigma_{red}$ = 0.68 & $K_{max}$ = 11\\
             &                          & N = 102(71)          &               \\
\hline
HCOOD        & $^{R}Q_{7}(J_{max}=42$)  & $\sigma$ = 60 kHz     & $J_{max}$ = 58\\
             & $^{Q}R_{K_{max=11}}$(28) & $\sigma_{red}$ = 0.66 & $K_{max}$ = 11\\
             &                          & N = 116(92)          &               \\
\hline

\end{tabular}
\vspace{1cm}
\caption{Branches with several transitions assigned and with the highest
quantum numbers are reported here. The notation $^{\Delta K_a}\Delta J_{K_a}(J_{low})$ was used. Other lines,
eventually with higher quantum numbers, were used in the fit and the highest $J$ and $K_a$ are reported.
The fit column gives the standard deviation, the reduced error and the numbers of lines used for the
global analysis. In brackets the numbers of lines assigned from our spectra. \label{tab:summary2}
}
\end{center}
\end{table}

\clearpage

\end{document}